\documentclass[%
superscriptaddress,
twocolumn,
notitlepage,
 amsmath,amssymb,
 aps,
]{revtex4-1}

\usepackage{graphicx}
\usepackage{dcolumn}
\usepackage{bm,color}

\begin{document}

\title{Multiply Charged Vortex States of Polariton
Condensates}

\author{Samuel N. Alperin}
\affiliation{Department of Applied Mathematics and Theoretical Physics, University of Cambridge, Cambridge CB3 0WA United Kingdom}
\author{Natalia G. Berloff}\email[correspondence address: ]{N.G.Berloff@damtp.cam.ac.uk}
\affiliation{Skolkovo Institute of Science and Technology Novaya St.,100, Skolkovo 143025, Russian Federation}
\affiliation{Department of Applied Mathematics and Theoretical Physics, University of Cambridge, Cambridge CB3 0WA United Kingdom}

\date{\today}
\newcommand{\red}[1]{\textcolor{black}{#1}}
\begin{abstract}
 The existence of quantized vortices is a key feature of Bose-Einstein condensates. In equilibrium condensates  only quantum vortices of unit topological charge are stable, due to the dynamical instabilities of multiply charged defects, unless supported by strong external rotation. Due to immense fundamental interest in the physics of these fundamental topological excitations, a great deal of work has been expended towards understanding ways to force their stability. Here we show that in photonic Bose-Einstein condensates of exciton-polariton quasiparticles pumped in an annular geometry, not only do the constant particle fluxes intrinsic to the system naturally stabilize multiply charged vortex states, but that such states can indeed form spontaneously during the condensate formation through a dynamical symmetry breaking mechanism. We elucidate the properties of these novel states, notably finding that they radiate acoustically, in a process analogous to the emission of gravitational waves from binary black holes. Finally, we show that the vorticity of these photonic fluids are fundamentally limited by a quantum Kelvin-Helmholtz instability, and therefore by the condensate radius and pumping intensity. This represents the first report of this instability - deeply fundamental in fluid dynamics - in a quantum photonic fluid.
\end{abstract}

\pacs{Valid PACS appear here}
\maketitle

From their macroscopic coherence it follows that Bose Einstein condensates (BECs) may only support rotational flow in the form of quantized vortices \cite{feynman1955chapter}. These vortices are thus topological in nature, and are characterized by a phase rotation of integer ($\ell$) steps of $2\pi$ around a phase singularity. However, while in principle quantized vortices may take on any topological charge, in practice it is understood that only vortices of charge $\ell=\pm 1$ are dynamically stable: higher order vortices quickly shatter into constellations of unit vortices due to energetics of the system. This shattering process has been detailed theoretically and observed experimentally in the context of stationary, harmonically trapped atomic BECs \cite{inouye2001observation,shin2004dynamical,engels2003observation}. The case is somewhat different for superharmonically trapped, rapidly rotated condensates, for which there exists a critical rotation rate above which the vorticity of the system becomes concentrated within a single effective core. These states, in which all vorticity is within a single effective core, has been called the \textit{giant vortex} state by its first experimental observers \cite{fischer2003vortex,fetter2005rapid}\footnote{We note that recently, similar giant vortex states have also been theorized to exist in the two-component quantum droplets \cite{li2018two}}. Such giant vortex is, therefore, different from  a state in which there is a single point singularity with topological charge magnitude greater than one -- \textit{multiply charged vortex}, however, in practice it is often impossible to distinguish between the two. On the one hand, the density in the vortex core is negligible, which hinders the resolution of singularity. On the other hand, the structure of interest is hydrodynamical, and thus only has meaning up to the length scales for which the hydrodynamical treatment applies. The classical field description being a long wavelength approximation of something which is in reality granular and nonclassical, the hydrodynamic description only applies down to the healing length. Singularities of like charge which are bound to within a healing length are thus, to any probe in the hydrodynamical regime, indiscernible from the theoretical \textit{multiply charged vortex}. Thus from here on we find it useful to call all such vortex structures \textit{multiply charged}.

In our study we focus on a BEC away from the thermodynamical equilibrium supported by continuous gain and dissipation such as polariton \cite{kasprzak2006bose}, photon \cite{klaers2010bose} or magnon \cite{demidov2008observation,nowik2012spatially} condensates.  To be more specific we use the example of polariton condensate however the results reported may be relevant to other nonequilibrium condensates. The exciton-polariton (polariton) is a bosonic quasiparticle composed of light (photons) and matter (excitons). Polaritons can be generated in optical semiconductor microcavities. In a typical experimental system, laser light is continuously pumped into the cavity to excite excitons (bound electron-hole pairs) in a semiconductor sample. The photons remain trapped in the cavity for some time, repeatedly being absorbed by the semiconductor to excite excitons, and then being re-emitted as the excitons decay. The excitons form superposition states (polaritons) with the photons, which behave neither as light or matter.  Due to the finite confinement times of the cavity photons, the polaritons in the condensate are themselves short lived. In this way the polariton condensate is fundamentally different from other condensates: here neither energy nor particle number need be conserved. Thus while a polariton condensate may settle into a \textit{steady state} (a state in which the wavefunction is time invariant up to a global phase shift), such a state is one in which dissipation is balanced by particle gain. The corollary is that steady state flows are possible. It is well understood that that the pattern forming capabilities of nonequilibrium, nonconservative systems is richer than those of equilibrium, conservative systems \cite{pismen2006patterns}, making the polariton condensate a fascinating object with which to explore the possibility of novel quantum hydrodynamical behaviors \cite{alperin2019formation}.

In this \red{article}, we show theoretically that multiply charged vortex states can appear spontaneously  and remain throughout the coherence time in a BEC of exciton-polariton quasiparticles excited by a ring-shaped laser profile, without the application of any external rotation, trapping potentials, or stirring. Previously, the spontaneous formation of multiply charged vortices of a given charge has been theoretically proposed and experimentally realised  in polariton condensates by pumping in an odd number of spots around a circle \cite{kalinin2017giant}, \red{or by the engineering of helical pumping geometries \cite{dall2014creation}. In the first case} the central vortex in this geometry is created driven by the antiferromagnetic coupling of the neighboring condensates and the frustration arising from their odd number. \red{In the second case the helical patterns are engineered so that the condensate is pumped explicitly with orbital angular momentum.} The multiply charged vortices we discuss in \red{here} differ from \red{these} works in the  geometry  considered (ring-pumped trapped condensates, long coherence times), formation mechanism (probabilistic and spontaneous during condensation) and the vortex properties (vortices exist on the maximum density background and so are truly nonlinear in nature). We describe their  formation, stability, and dynamics.  The dynamics of two and more interacting multiply charged vortices are also studied. We find that our results apply for a wide range of possible experimental parameters, suggesting that these structures are general to ring-pumped trapped polariton BECs \red{in the strong coupling regime}.


The dynamics of the polariton BEC in the mean field are described by the complex Ginzburg-Landau equation (cGLE) coupled to a real reservoir equation representing the bath of hot excitons in the sample, nonresonantly excited by the spatially resolved laser pump profile $P({\bf r})$ \cite{carusotto2013quantum,wouters2007excitations,keeling2008spontaneous}

\begin{figure}[t]
\centering
\includegraphics[width=\columnwidth]{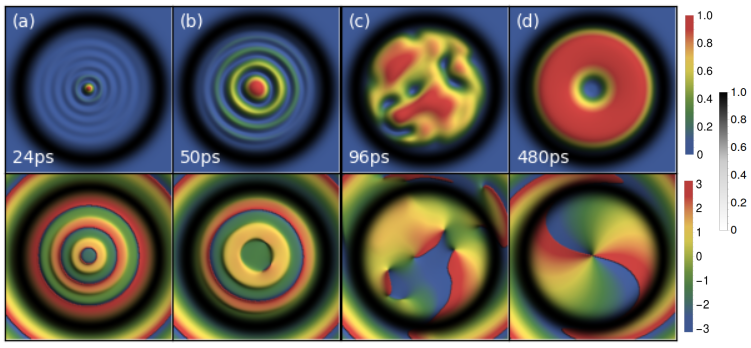}
 \caption{Spontaneous formation of a multiply charged quantum vortex in a ring pumped polariton condensate by numerical integration of Eqs.~(\ref{GPE}-\ref{NR}). Density (top row) and phase (bottom row) snapshots are shown at various stages of the condensate formation. \red{For clarity, each density profile is rescaled to unit maxima. The pumping profiles are superposed in black (in units of P), showing the spatial separation between the pump and the condensate.} At the beginning of the condensate formation; due to the pump geometry, matter wave interference leads to annular zeros in the wavefunction (a). These ring singularities are unstable to dynamical instability, \red{become asymmetrical} (b) and can be observed to break into more stable unit vortices in as the condensate continues to develop. The condensate fills a disk shaped region with near uniformity within the ring pump, but remaining vortices interact chaotically in (c). The vortex turbulence eventually decays, leaving a net topological charge \cite{gladilin2019multivortex,gladilin2017interaction}. Repeating these simulations with different random initial conditions, the magnitude and sign of the final vorticity varies. Here $P=5$ and $r_0=11.5\mu m$.}
  \label{spont}
\end{figure}

\begin{figure}[b!]
\includegraphics[width=\columnwidth]{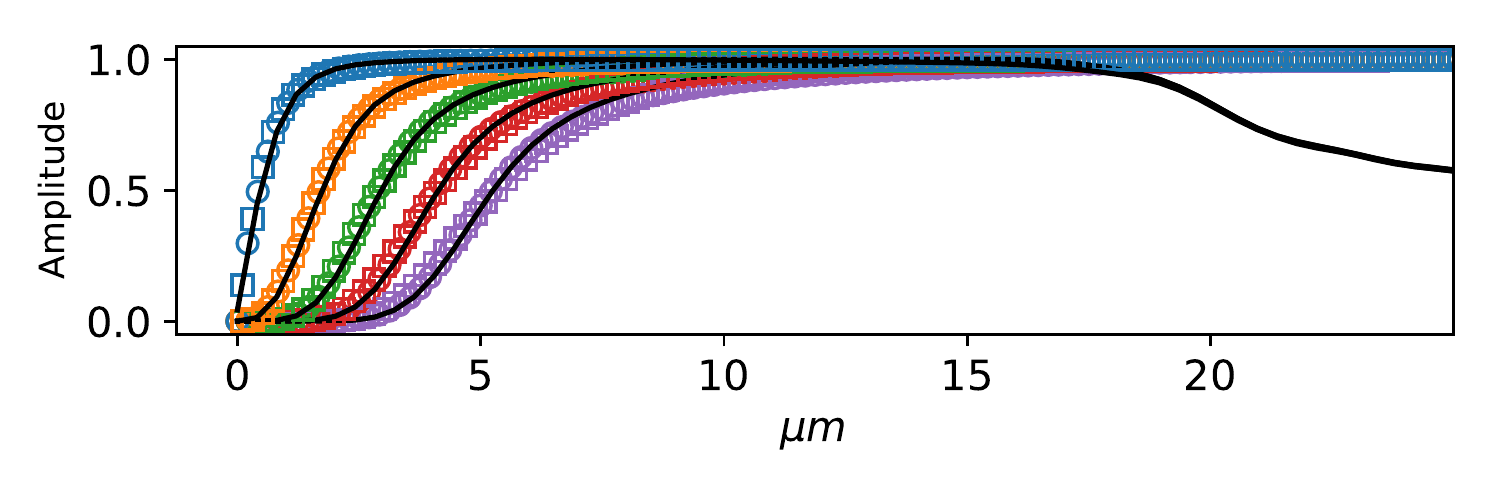}
\caption{Wavefunction amplitude cross sections $\sqrt{\rho(r)}$ of multiply charged vortices. For clarity, and without loss of generality, we show only odd topological charges less than $\ell=10$. Profiles from the full numerical integration of Eqs.~(\ref{GPE}-\ref{NR}) (normalized) for  $r_0=20\mu m$, $P=12$, and $\gamma=0.3$ are shown in black, and illustrate the decay of the condensate near the pump ring. The numerical solutions of the reduced equation Eq.~(\ref{tildeA}) are marked by circles colored by charge, and the corresponding fits to the ansatz Eq.~(\ref{fit}) by squares with matching colors. From these fits, we can write the approximate parameterization of Eq.~(\ref{tildeA}) as $n(\ell)=(1.1)\ell^{1.6}-2.8$ and $w(\ell)=2.3+(0.6)ln(\ell)$).}
  \label{profile}
\end{figure}
\begin{eqnarray}
  \label{GPE_dim}
 i \hbar \partial_t\psi &= &
 -\frac{\hbar^2}{2m}(1-i\hat{\eta} N_R) \nabla^2 \psi
    + U_0|\psi|^2 \psi+ g_R N_R \psi  \\ \nonumber &+&\frac{i\hbar}{2}(R_R N_{R}- \gamma_C)\psi\\
 \partial_t N_R&=&P-(\gamma_R+R_R|\psi|^2)N_R \label{NR_dim}
 \end{eqnarray}
 in which $\psi$ represents the condensate wavefunction, $N_R$ and the exciton reservoir density. $U_0$ and $g_R$ give the polariton-polariton and exciton-polariton interaction strengths, $R_R$ and $\hat{\eta}$ represent the scattering and diffusion rates. The effective mass of the polariton is given by $m$. Finally, the loss rates of excitons and polaritons are described by $\gamma_C$ and $\gamma_R$. To rewrite these equations in a more amenable, nondimensional form, we employ the transforms $\psi \rightarrow \sqrt{\hbar R_R /2 U_0 l^2_0}\psi$,$t\rightarrow 2l^2_0t/ \hbar R_R$, $r\rightarrow\sqrt{\hbar l^2_0/(mR_R)}r$,$N_R\rightarrow N_R/l_0^2$,$P\rightarrow R_R P/2 \hbar l^2_0$, and we define the nondimensionless parameters $g=2g_R/R_R$, $b_0=2\gamma_R l_0^2/\hbar R_R$, $b_1= R_R/U_0$,$\eta=\hat{\eta}/l^2_0$,$\gamma=\gamma_C l^2_0/R_R$, and $\gamma=\gamma_Cl^2_0/R_R$, where we set $l_0=1\mu m$. This yields \cite{kalinin2018simulating}

 
\begin{eqnarray}
  \label{GPE}
 i \partial_t\psi &= &
 -(1-i\eta N_R) \nabla^2 \psi
    + |\psi|^2 \psi+g N_R \psi \\ \nonumber &+& i(N_{R}-\gamma)\psi \label{Psi}\\
 \partial_t N_R&=&P-(b_0+b_1|\psi|^2)N_R, \label{NR}
 \end{eqnarray}

Polaritons  can be confined all-optically by shaping the excitation laser beam. By using spatial light modulators to shape the optical excitation, ring shaped confinements were generated with condensates forming inside the ring \cite{cristofolini2013optical, askitopoulos2013polariton}, \red{and have been predicted to support the spontaneous formation of unit vortices \cite{yulin2016spontaneous}}. Long-lifetime polaritons in  ring traps are emerging as a platform for studies of fundamental properties of polariton condensation largely decoupled from the excitonic reservoir and, therefore, having significantly larger coherence times \cite{sun2017bose, askitopoulos2016nonresonant, pickup2018optical, alyatkin2019optical}.  We represent the profile of the ring pump by a Gaussian annulus of the form $P(r,\theta,t)=P e^{-\alpha (r-r_0)^2}$ \red{with inverse width $\alpha$ and radius $r_0$}, which excites local quasiparticles which then flow outward. The closed-loop pump geometry  has two major implications. The first is that the condensation threshold is first achieved not where the sample is pumped, but \textit{within} the borders of the pumping ring. This results in the \red{effective} spatial separation of Eqs.~(\ref{GPE}-\ref{NR}), which makes the parameters related to the excitonic reservoirs such as $b_0$, $b_1$, and $g$ irrelevant to the condensate dynamics up to a change of pump strength. The second and most critical implication of the ring pump geometry  is the existence of constant  fluxes towards the centre of the ring. Such fluxes carry the matter together with spontaneously formed vortices and force vortices to coalesce. 

\begin{figure}
\includegraphics[width=\columnwidth]{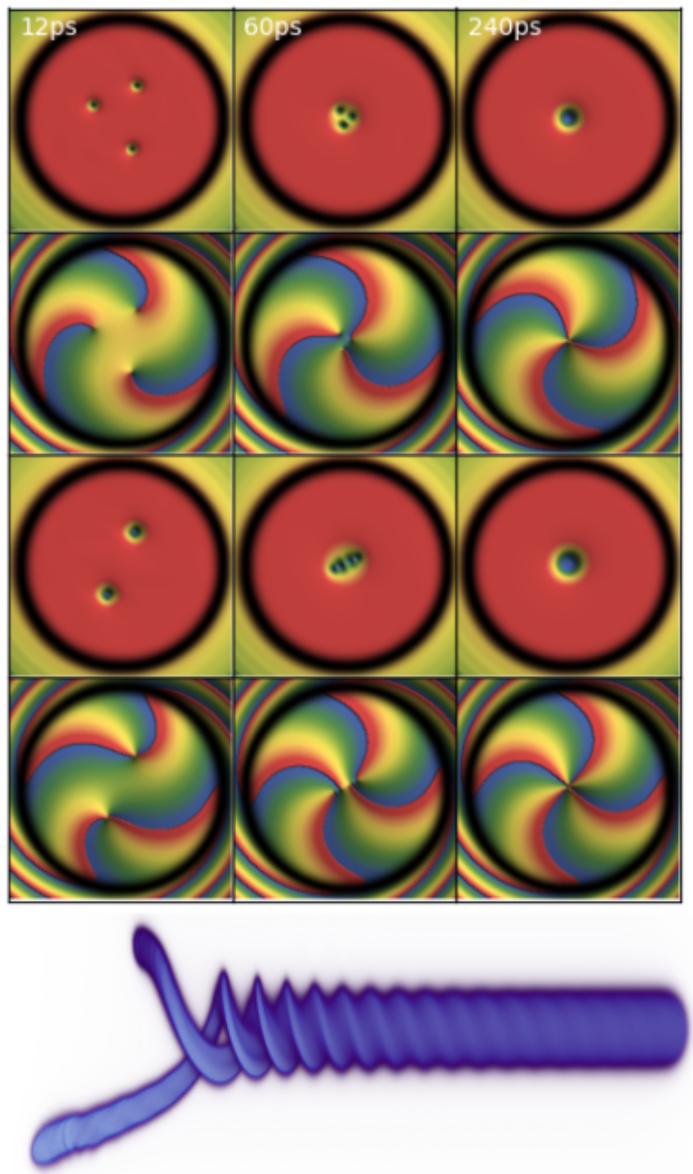}
 \caption{From Top: Density (first row) and phase (second row) resolved dynamics of three unit vortices of like charge in a condensate formed within the boundary of an annular pump (green). Over time, the three vortices approach each other in an inward spiral, eventually merging to an inter-singularity length scale less than the healing length of the condensate. Density (3rd row) and phase (4th row) of two doubly charged vortices (each having topological charge $n=2$), which over time merge into a single fourth-order vortex. Here $P=10$ and $r_0=15\mu m$. \red{Colour scales are the same as in Fig. \ref{spont}.} At bottom, a density isosurface of the two merging second-order vortices, with time shown along the horizontal axis \red{, from 12 ps (left) to 240 ps (right).}}
  \label{coel}
\end{figure}

It is well known that vortices can form during the rapid condensation of a Bose gas, via the Kibble-Zurek mechanism \cite{kibble1976topology,kibble1980some,zurek1985cosmological,zurek1996cosmological,damski2010soliton,bauerle1996laboratory,ruutu1996vortex,bowick1994cosmological,maniv2003observation,weiler2008spontaneous}. However, in our system there exist  a different mechanism of spontaneous defect generation in our system, which requires a relatively \textit{slow} condensate formation. Due to the inward flow of particles in our system, the condensation threshold is reached first in the center of the system. Assuming a large enough ratio of new particle flow to dissipation, this young condensate will grow into a relatively uniform disk within the boundary of the pump. However, in between these two stages, radial matter wave interference is to be expected, with higher frequency during the early stages of condensation. The zeros of the radial interference pattern are well studied under a different name: the dark ring soliton \cite{staliunas2003transverse,pismen2006patterns,aranson2002world}. As has been shown previously, these dark solitons are unstable to transverse (`snake') perturbations, and break apart into pairs of unit vortices of opposite charge \cite{carr2006vortices,kevrekidis2007emergent}. Thus for a slowly condensing system, it is reasonable to expect that these solitons have enough time to break down to produce a chaotic array of vortex singularities. This process resembles a two-dimensional case of the collapsing bubble mechanism of vortex nucleation \cite{berloff2004vortex}.
As the condensation process completes and the vortex turbulence decays, there is some finite chance of the condensate being left with a net topological charge, as vortex pairs may unbind near the boundary and one or the other may leave to annihilate with its image. These like-charged vortices would then coalesce in the center of the condensate.

 Direct numerical integration of Eqs.~(\ref{GPE}-\ref{NR}) not only confirms that this process can take place, but that for low pump power, the condensate takes on a net topological charge more often than not \footnote{Fourth order Runga-Kutta integration is used. The initial wavefunction is set to a profile of low amplitude random noise. All simulations are repeated for many of these profiles.}. We reiterate that this coalescence of vortices is despite the lack of external rotation, \red{or sample nonuniformity}. Repeating the numerical experiment with many iterations of random initial wavefunction noise, we find multiply charged vortex states of stochastic sign and magnitude. The average topological charge magnitude is found to depend significantly on the radius of the pump ring, increasing for larger radii. An example of these dynamics is presented in Fig. \ref{spont}, which shows the main steps in the process by which the condensate spontaneously adopts a topological charge of two: the formation of a central condensate surrounded by annular discontinuities in Fig. \ref{spont}(a), the breakdown of an annular discontinuity into vortex pairs in Fig. \ref{spont}(b), vortex turbulence in Fig. \ref{spont}(c), and the final bound vortex state Fig. \ref{spont}(d). For Fig. \ref{spont} we use the system parameters $\eta=0.3$, $\gamma=0.05$, $g=1$, $b_0=1$, $b_1=6$, but the result was found not to depend sensitively on these choices; up to a rescaling of pump strength this behavior was reconfirmed for a large range of sample parameters: $g \in [0.1-2]$, $b_0 \in [0.01-10]$, for $\gamma \in [0.05-0.1]$, and for all reasonably physical values of $\eta$ (including $\eta=0$.)
 

An advantage \red{ of the spatial separation of} the condensate from the reservoir in ring pumped geometry is in the enhanced   coherence time that exceeds the individual particle lifetime by three orders of magnitude \cite{askitopoulos2019giant}. Therefore, spontaneously created multiply charged vortices can be observed in single shot experiments within one condensate realisation. Another way to study the multiply charged vortices is to imprint  them explicitly upon a fully formed, uniform condensate \cite{amo2009collective}.
This allows for the study of the structure and dynamics of carefully controlled systems of vortices. To model the result of experimental pulsed phase imprinting, we first model the formation of fully developed non-singular condensate disks. To prevent the spontaneous formation of vortices by the process described above, a relatively strong pump amplitude is used, so that the condensate forms too quickly for the decay of ring-singularities into vortices. After the background condensate is formed, phase singularities are imprinted instantaneously and their dynamics is observed. To first understand the structure of isolated multiply charged vortices, we imprint a series of condensates with different topological charges, and allow these structures to form steady states. When imprinted in equilibrium BEC, multiply charged vortices quickly break into vortices of a single unit of quantization \cite{kawaguchi2004splitting}.

From the spatial separation of the condensate and the reservoir, the reservoir density is negligible near the central core of the multiply charged vortex, so that Eq.~(\ref{GPE}) takes the familiar form of the damped nonlinear Schr\"odinger equation (dNLSE):
$
i \partial_t \psi=-\nabla^2\psi+|\psi|^2\psi -i\gamma \psi.
$

Under the Madelung transformation $\psi=\mathcal{A}\exp[iS-i \mu t]$ where $\mu$ is the chemical potential, the velocity is  the gradient of the phase $S$: $\textbf{u}=\nabla S$ and the density  is $\rho(r)=\mathcal{A}^2,$ the imaginary part of the dNLSE yields
$\nabla \cdot (\rho \textbf{u})=-\gamma \rho.$ Except for a narrow spatial region where the density heals itself from zero to the density of the vortex free state the density is almost a constant, so the radial component of the velocity becomes $u_r=-\gamma r$. The real part of the dNLSE reads 
\begin{equation}
\partial_r^2\mathcal{A}+\partial_r\mathcal{A}/r+(\mu- \textbf{u}^2 -\mathcal{A}^2)\mathcal{A}=0, \label{A}
\end{equation}
which coincides with the corresponding steady state equation for the equilibrium condensates where  velocity profile   plays the role of the external potential. We therefore expect the structure of the vortices to be similar to those in equilibrium condensates with the external potential given by $\textbf{u}^2$. Close to the centre of the condensate the velocity becomes $\textbf{u}=-\gamma r \hat{r}+\frac{\ell}{r}\hat{\theta},$
\begin{figure}
\includegraphics[width=\columnwidth]{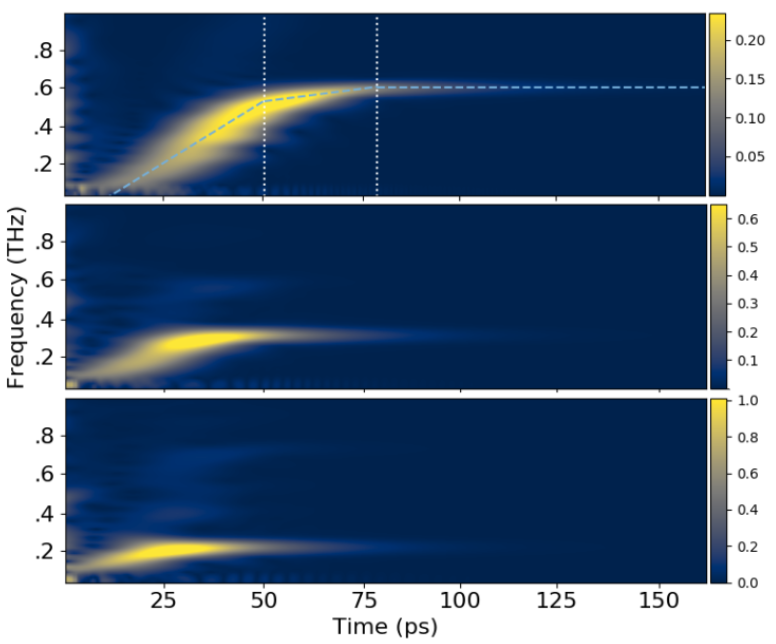}
 \caption{\red{Power spectral density of acoustic waves radiated by the approach and merger of two vortices, resolved in time-frequency space. Top panel shows the} merger of two unit vortices, and vertical lines mark the time of transition from well separated vortices to vortices sharing a common low-density core (left) and \red{the time at which the singularities have merged to within a healing length} (right).  Middle and bottom panels show the acoustic spectra of two merging doubly charged vortices (middle) and two merging triply charged vortices (bottom). The ring pump radius is $20 \mu m$. \red{In the case of two single vortices, one vortex is imprinted at the condensate center, and the other at a distance of $18 \mu m$ from the center. In the cases of two multiply charged vortices, both vortices are imprinted $18 \mu m$ from the center.}}
  \label{freq}
\end{figure} where $\hat{r}$ and $\hat{\theta}$ are unit vectors in polar coordinates. When this expression for $\textbf{u}$ is substituted into Eq.~(\ref{A}) it becomes the equation on the vortex amplitude  in the centre of the harmonic trap, where $\gamma$ characterises the frequency of the ``trap."  In the vortex core, for small $r$, the centrifugal velocity dominates the radial velocity, so the equation on the rescaled amplitude $A =\mathcal{A}/\sqrt{\mu}$ with $\tilde r =\sqrt{\mu}r$ becomes 
\begin{equation}
\partial_{\tilde r^2}{A}+\partial_{\tilde r}{A}/{\tilde r}+\biggl(1-\frac{\ell^2}{\tilde r^2}-{A}^2\biggr){A}=0. \label{tildeA}
\end{equation}
The profiles $A$  take the approximate form 
\begin{equation}
A=\frac{\tilde r^{|\ell|}}{(\tilde r^{n}+w)^{|\ell|/ n}},
\label{fit}
\end{equation}
with parameters $w$ and $n$ in which we incorporated the power expansion behaviour of the amplitude $A\sim \tilde r^{|\ell|}$ as $\tilde r\rightarrow 0$.  Figure \ref{profile} shows the amplitude cross-section profiles of stable giant vortices with different topological charges $\ell \in \{1,2,...,10\}$ as the solutions of Eq.~(\ref{tildeA}), along with  Eq. (\ref{fit}).

\begin{figure}
\includegraphics[width=\columnwidth]{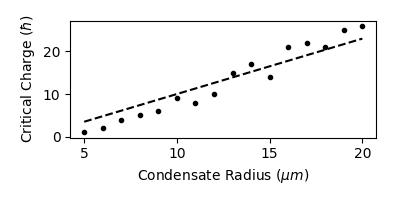}
 \caption{Plotted is the critical topological charge at which the KHI sets in, as a function of pump radius. This is obtained by the  direct numerical simulations of Eqs.~(\ref{GPE}-\ref{NR}). The dashed line represents the theoretical expectation of Eq. \ref{kh_eq}.
 }
  \label{kelvin}
\end{figure}

Next we consider the  arrangements of multiple \red{multiply charged} vortices imprinted away from the trap center and brought together by the radial fluxes. Fig. \ref{coel} shows two examples of the coalescence dynamics of imprinted phase defects. In the first case, three unit vortices coalesce while moving in  inward spirals towards the center of the condensate, where there is no net lateral flow. In the second case, which shows the coalescence of two doubly charged vortices, it is observed that both doubly charged vortices hold together for a while before  merging in the center to form a single vortex of multiplicity four. These results are found to be repeatable for a wide range of system parameters, suggesting that this behaviour is to be expected for any system parameters which allow the formation of the trapped condensate within a ring pump.  \red{We note that in this system, the center of the condensate corresponds to the location of maximum background fluid density, in stark contrast to systems designed to collect virtual vorticity into a low-density area \cite{cozzini2006vortex}.}

The process of coalescence of two or more vortices is interesting from the point of view of analog gravity and ``cosmology in the lab" and the analogies can be drawn to the collision of  two or more \red{black holes}  \cite{abbott2017gw170817,unruh1981experimental, volovik2003universe,steinhauer2014observation}. As two (or more) vortices merge while spiralling around the center  they excite density waves in the otherwise uniform background fluid. These acoustic excitations are long lived, and take on a frequency set by the angular frequency of the vortex spiral. For well separated vortices, this frequency increases consistently as time progresses and their separation shrinks. However, as the vortices begin to share a these dynamics become even more complicated and the new physics dominated by the processes in the vortex core emerges \cite{berloff2009turbulent}.

\begin{figure}
\includegraphics[width=\columnwidth]{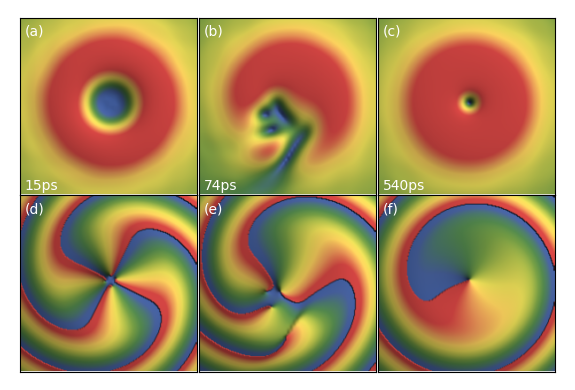}
 \caption{For a great enough topological charge (compared to the size of the condensate), the rotational flow at the boundary of the condensate reaches the critical velocity for the Kelvin-Helmholtz instability to set in, which results in the reduction of topological charge via the nucleation of new vortices, with charges opposite to that of the \red{multiply charged} vortex and further pair annihilation. Shown are density profiles from a direct simulation of Eqs. Eqs.~(\ref{GPE}-\ref{NR}) exhibiting this process (radius $7 \mu m$,$P=10$). The initial topological charge is imprinted one quanta at a time, and the dynamics observed. After the fourth quanta of rotation is imprinted, the system loses stability and expels some rotation through the KHI mechanism, ending with unit topological charge.
 }
  \label{kelvin_dyn}
\end{figure}

Figure \ref{freq} shows the relative amplitudes of the density waves radiated during the motion of two vortices of unit charge imprinted with a large initial separation. The average frequency of acoustic radiation is found to increase with time at a fixed rate until the \red{vortex cores begin to overlap} (left vertical line). During this phase, the frequency distribution narrows and the average radiation frequency increases linearly at a much lower rate than in the well-separated vortex regime. This continues until  \red{the singularities within the core overlap within a healing length} (right vertical line), after which a fixed narrow band of acoustic radiation is emitted. These acoustic excitations are not unlike the gravitational waves found in the case of binary astronomical systems, which also exhibit a three stage signature in their coalescence \cite{abbott2017gw170817}.

\red{ Of course, multiply charged vortices may also collide: we find that from merger of two equal} multiply charged vortices of increasing topological charge, the characteristic acoustic resonances have decreasing frequency, in the near-terahertz regime. \red{This is because the effective mass of the vortex increases with topological charge, so that vortices of larger multiplicity orbit more slowly. As expected, we see that in contrast, when multiple singly charged vortices placed evenly about a common radius merge, they emit higher frequency radiation as the number of unit vortices is increased.} \red{Once the multiply charged vortex has formed and is allowed to settle, low energy density perturbations can be applied to the condensate. To model this, we simulate the effect of a small Gaussian laser pump pulse centered on the vortex. The observed effect is the emission of  an acoustic energy pulse at the characteristic frequency of the vortex, as is seen in from the merger of the equivalent number of unit vortices. As in any physical system there exist many small perturbations due to intrinsic disorder, it is likely that multiply charged vortices in a real system are regularly being excited and emitting acoustic radiation.}

{\it Kelvin-Helmholtz instability.} Next we will establish the limit on the vortex multiplicity that the trapped condensate can support. This limit is set by the maximum counterflow velocity that can be supported between the condensate and the reservoir, therefore, is determined by  the onset of a Kelvin-Helmholtz instability (KHI). KHI is the dynamical instability at the interface of two fluids when the counterflow velocity exceeds a criticality. It appears in variety of disparate systems, both  classical and quantum, but has never been discussed in the context of the polaritonic systems. In quantum fluids  KHI manifests itself via nucleation of vortices at a counterflow velocity  exceeding the local speed of sound $v_c=\sqrt{\frac{U_0\rho}{m}}$. It has been extensively studied  for the interface between different phases of $^3$He \cite{volovik2002kelvin}, two components in atomic BECs \cite{takeuchi2010quantum} or for the relative motion of superfluid and normal components of $^4$He \cite{hanninen2014vortex, lushnikov2018exact}. In trapped condensates considered here, the counterflow is that between the condensate of radius $R$ (which rotates with velocity $\frac{\hbar}{m}\frac{|\ell|}{ R}$ \red{at the boundary}), and the reservoir particles along the ring, that are stationary. Thus it is expected that KHI should be initiated when the topological charge of the multiply charged vortex state is high enough so that the \red{velocity of} condensate \red{particles} at the ring pump radius reach $v_c$. Thus the maximum topological charge $\ell_c$ allowed is set by 
\begin{equation}
    |\ell_{c}|=\frac{ \sqrt{R^2 mU_0 \rho}}{\hbar}.
\label{kh_eq}
\end{equation}
Fig. \ref{kelvin} shows Eq. \ref{kh_eq} (dashed line) along with the critical topological charges found \red{by} direct numerical integration of Eqs.~(\ref{GPE}-\ref{NR}) (dots). In these numerical experiments, we begin with a fully developed, vortex-free condensate. A unit topological charge is imprinted in the center of the condensate, and the system is allowed to settle, before another unit charge is added. This process is repeated until the onset of the KHI leads to the vortex nucleation followed by annihilation of vortex pairs and, therefore, by the  reduction in the topological charge of the system. This dynamical process is shown in Fig. \ref{kelvin_dyn}.

In conclusion, we have shown that exciton-polariton condensates excited by an annular pump can spontaneously rotate despite \red{a uniform sample} and no angular momentum applied, forming multiply charged vortices. The formation, dynamics and structure of these vortices were studied. 

\textbf{Disclosures.} The authors declare no conflicts of interest.

\end{document}